\documentclass{aa}
\usepackage{graphicx,bm,natbib,url,txfonts,color}
\newcommand{\EQ}{\begin{equation}}
\newcommand{\EN}{\end{equation}}
\newcommand{\EQA}{\begin{eqnarray}}
\newcommand{\ENA}{\end{eqnarray}}

\newcommand{\EEq}[1]{Equation~(\ref{#1})}
\newcommand{\Eq}[1]{Eq.~(\ref{#1})}
\newcommand{\Eqs}[2]{Eqs.~(\ref{#1}) and~(\ref{#2})}

\newcommand{\eqs}[2]{(\ref{#1}) and~(\ref{#2})}
\newcommand{\App}[1]{Appendix~\ref{#1}}

\newcommand{\Fig}[1]{Fig.~\ref{#1}}

\newcommand{\Figp}[2]{Fig.~\ref{#1}({#2})}

\newcommand{\Tab}[1]{Table~\ref{#1}}

{}
{}
{}

{}
{}
{}
{}
{}
{}
{}
{}
{}
{}
{}
{}
{}
{}
{}
{}
{}
\newcommand{\meanUU}{\overline{\bm{U}}}

{}

\newcommand{\meanA}{\overline{A}}
\newcommand{\meanB}{\overline{B}}

\newcommand{\hatBB}{\hat{\bm{B}}}
{}

{}
{}
{}

\newcommand{\meanAA}{{\overline{\bm{A}}}}
\newcommand{\meanBB}{{\overline{\bm{B}}}}
\newcommand{\meanJJ}{{\overline{\bm{J}}}}

\newcommand{\bb}{\bm{b}}

\newcommand{\uu}{\bm{u}}

\newcommand{\SSS}{\mbox{\boldmath $S$} {}}

\newcommand{\nab}{{\bm{\nabla}}}

\newcommand{\ii}{{\rm i}}

\newcommand{\sgn}{{\rm sgn}  \, {}}

\newcommand{\dd}{{\rm d} {}}

\def\Rey{\mbox{\rm Re}}
\def\Imag{\mbox{\rm Im}}

\def\etat{\eta_{\rm t}}

\def\etaT{\eta_{\rm T}}

\def\tautd{\tau_{\rm td}}

\def\half{{\textstyle\frac{1}{2}}}
\newcommand{\yapj}[3]{ #1, {ApJ,} {#2}, #3}

\newcommand{\yapjl}[3]{ #1, {ApJ,} {#2}, #3}
\newcommand{\yapjs}[3]{ #1, {ApJS,} {#2}, #3}

\newcommand{\yan}[3]{ #1, {Astron.\ Nachr.,} {#2}, #3}

\newcommand{\ymhdn}[3]{ #1, {Magnetohydrodyn.} {#2}, #3}
\newcommand{\yana}[3]{ #1, {A\&A,} {#2}, #3}

\newcommand{\ygafd}[3]{ #1, {Geophys.\ Astrophys.\ Fluid Dyn.,} {#2}, #3}

\newcommand{\yjfm}[3]{ #1, {J.\ Fluid Mech.,} {#2}, #3}

\newcommand{\ypp}[3]{ #1, {Phys.\ Plasmas,} {#2}, #3}
\newcommand{\ysov}[3]{ #1, {Sov.\ Astron.,} {#2}, #3}

\newcommand{\yphyd}[3]{ #1, {Physica D,} {#2}, #3}

\newcommand{\yprsa}[3]{ #1, {Proc.\ Roy.\ Soc.\ Lond. A,} {#2}, #3}

\newcommand{\ymn}[3]{ #1, {MNRAS,} {#2}, #3}

\newcommand{\ypre}[3]{ #1, {Phys.\ Rev.\ E,} {#2}, #3}

\newcommand{\yjcp}[3]{ #1, {J.\ Comput.\ Phys.,} {#2}, #3}
\newcommand{\yjour}[4]{ #1, {#2}, {#3}, #4}

\newcommand{\ybook}[3]{ #1, {#2} (#3)}
\newcommand{\yproc}[5]{ #1, in {#3}, ed.\ #4 (#5), #2}

\newcommand{\dmn}[2]{ #1, {MNRAS}, DOI:#2}
\def\apjs{ApJS}

\AASection{Astron. Astrophys. 598, A117 (2017)}

\title{Analytic solution of an oscillatory migratory $\alpha^2$ stellar dynamo}
\author{A. Brandenburg\inst{1,2,3,4}}
\institute{
Nordita, KTH Royal Institute of Technology and Stockholm University, 10691 Stockholm, Sweden
\and
JILA and Department of Astrophysical and Planetary Sciences, University of Colorado, Boulder, CO 80303, USA
\and
Laboratory for Atmospheric and Space Physics, University of Colorado, Boulder, CO 80303, USA
\and
Department of Astronomy, Stockholm University, 10691 Stockholm, Sweden
}
\date{Received November 8, 2016, accepted December 1, 2016, $ $Revision: 1.85 $ $}

\begin{document}

\abstract{
Analytic solutions of the mean-field induction equation predict
a nonoscillatory dynamo for homogeneous helical turbulence or constant
$\alpha$ effect in unbounded or periodic domains.
Oscillatory dynamos are generally thought impossible for constant $\alpha$.
}{
We present an analytic solution for a one-dimensional bounded
domain resulting in oscillatory solutions for constant $\alpha$,
but different (Dirichlet and von Neumann or perfect conductor
and vacuum) boundary conditions on the two boundaries.
}{
We solve a second order complex equation and superimpose two
independent solutions to obey both boundary conditions.
}{
The solution has time-independent energy density.
On one end where the function value vanishes, the second derivative is finite,
which would not be correctly reproduced with sine-like expansion functions
where a node coincides with an inflection point.
The field always migrates away from the perfect conductor boundary toward
the vacuum boundary, independently of the sign of $\alpha$.
}{
The obtained solution may serve as a benchmark for numerical dynamo
experiments and as a pedagogical illustration that oscillatory migratory
dynamos are possible with constant $\alpha$.
}

\maketitle

\section{Introduction}

The magnetic fields in stars and galaxies are believed to be generated
and maintained by large-scale dynamos that convert kinetic energy
into magnetic energy through an inverse cascade \citep{PFL76}.
With the development of mean-field theory \citep{Par55,SKR66},
this complicated three-dimensional process became amenable to simpler
analytic and numerical treatments in one and two dimensions.

The best known mean-field effect is the $\alpha$ effect, which emerges
from the parameterization of the turbulent electromotive force in terms of
the mean field in the form
\begin{equation}
\overline{\uu\times\bb}=\alpha\meanBB-\etat\nab\times\meanBB,
\end{equation}
where $\uu$ and $\bb$ are the fluctuating velocity and magnetic fields,
overbars denote averaging, and $\meanBB$ is the mean magnetic field.
Here, $\alpha$ quantifies the $\alpha$ effect and $\etat$ is the turbulent
magnetic diffusivity.
Both are in principle functions of position, but in the present paper
we will treat them as constants.

The earliest model of a dynamo for the Sun goes back to \cite{Par55},
who considered the additional presence of differential rotation, which
is referred to as the $\Omega$ affect.
In the presence of both $\alpha$ and $\Omega$ effects, there are self-excited
oscillatory plain wave solutions in unbounded domains.
They take the form of traveling waves \citep{Par55}.
Specifically, if $\alpha$ is positive in the north and negative in the
south, and the differential rotation has a negative radial gradient,
waves are traveling equatorward, providing thus an explanation for the shape
of Maunder's butterfly diagram \citep{Mau04}.
The first global axisymmetric two-dimensional models of such dynamos go back
to the seminal work of \cite{SK69a}.
These dynamos are referred to as $\alpha\Omega$ dynamos.

In the absence of differential rotation, a plain wave solution ansatz
leads to non-oscillatory dynamos if $\alpha$ exceeds a
certain threshold ($\alpha>\etat k$, where $k$ is the wavenumber).
Such dynamos are referred to as $\alpha^2$ dynamos.
The dynamo of the Earth is believed to be an example of an $\alpha^2$
dynamo, because shear is expected to be weak.
Axisymmetric models of dynamos of this type where presented by
\cite{SK69b}.
The non-oscillatory property of such dynamos is consistent with the
noncyclic nature of the Earth's magnetic field.
In galaxies, on the other hand, shear is important, so they are examples
of $\alpha\Omega$ dynamos.
However, asymptotic solutions have shown that such dynamos are
non-oscillatory owing to the flat geometry in which such dynamos are
embedded \citep{VR71}.

Numerical investigations of $\alpha^2$ dynamos revealed only
nonoscillatory solutions \citep{Rae80}, until \cite{SSR85} found that,
under certain conditions, oscillatory solutions are here possible, too.
They associated this with the non-selfadjointness of the problem.
In fact, the possibility of oscillatory solutions to an $\alpha^2$ dynamo
was already mentioned earlier by \cite{RSS80} in a study of disk dynamos
with a strongly localized $\alpha$ effect.
In 1987, there appeared two back-to-back papers that demonstrated
conclusively that $\alpha^2$ dynamos can in principle be oscillatory
provided the $\alpha$ effect is non-constant \citep{BS87,RB87}.
This possibility remained mainly an academic curiosity without real
astrophysical interest at the time.

In subsequent years, attention was drawn to the possibility that
global dynamos with radially dependent $\alpha$ can exhibit oscillatory
solutions \citep{SG03}.
Meanwhile, direct numerical simulations of helically forced turbulence
have shown a strong similarity between $\alpha$ effect dynamos and
turbulent three-dimensional dynamos with fluctuating magnetic fields and
nonvanishing mean fields.
These dynamos turned out to be equivalent to those predicted from
$\alpha$-effect dynamos \citep{B01}.
\cite{MTBM09} applied such dynamos to spherical wedges with
helically forced turbulence.
When the helicity of the forcing was assumed such that it changes sign
about the equator, \cite{MTKB10} found oscillatory solutions with
equatorward migration similar to what occurs in the Sun.
\cite{KMCWB13} argued that such an effect can explain the equatorward
migration in their spherical wedge-geometry dynamos, even though shear
was still present and, as it turned out later, responsible for an
$\alpha\Omega$-type dynamo in this case \citep{WKKB14}.
In other simulations, however, the argument in favor of an $\alpha^2$ dynamo
could still be supported \citep{MS14}.

Corresponding mean-field solutions were presented by \cite{BCC09} for
dynamos in Cartesian geometry with $\alpha$ profiles proportional to $z$.
\cite{CBKK16} showed that such dynamos are not necessarily expected
to operate in spherical shells that extend all the way to the poles,
unless the turbulent magnetic diffusivity becomes small at high latitudes.
The true applicability of such $\alpha^2$ dynamos to stars remains
therefore questionable.
Nevertheless, such dynamos are gaining in importance in view of the
many numerical studies of turbulent dynamos, in which the helicity profile
is non-uniform \citep{MBKR14,JBMKR16} and/or the boundary conditions
on the two sides of the domain are different \citep{JBKR16}.
This has led to the possibility that oscillatory $\alpha^2$ dynamos
might actually be possible for constant $\alpha$, provided the boundary
conditions are indeed different and the two sides.
If this is the case, it should be possible to construct exact analytical
solutions of such an oscillatory migratory $\alpha^2$ dynamos.
The purpose of the present paper is therefore to present such a solution.
The fact that such a solution can be obtained analytically is significant
not only as a benchmark for numerical studies, but also as a clear
textbook-style demonstration of oscillatory $\alpha^2$ dynamos.

\section{Statement of the problem}

The equation for an $\alpha^2$ dynamo with total (sum of microphysical
and turbulent) magnetic diffusivity, $\etaT=\eta+\etat$, is given by
\begin{equation}
\frac{\partial\meanAA}{\partial t}=\alpha\nab\times\meanAA
-\etaT\nab\times\nab\times\meanAA,
\label{induct}
\end{equation}
where $\meanAA$ is the mean magnetic vector potential in the Weyl gauge,
and the mean magnetic field is $\meanBB=\nab\times\meanAA$.
We nondimensionalize by measuring lengths in units of $k_1^{-1}$,
where $k_1$ is the wavenumber of the most slowly decaying mode,
and time is measured in units of the turbulent--diffusive time,
$\tautd=(\etaT k_1^2)^{-1}$.
Velocities are measured in units of $\etaT k_1$, so in the following
we denote by $\alpha$ the nondimensional $\alpha$ effect,
$\alpha/\etaT k_1$.
We now consider a one-dimensional domain, so the governing equations are,
\begin{equation}
\frac{\partial\meanA_x}{\partial t}=-\alpha\frac{\partial\meanA_y}{\partial z}
+\frac{\partial^2\meanA_x}{\partial z^2},
\label{eq1}
\end{equation}
\begin{equation}
\frac{\partial\meanA_y}{\partial t}=+\alpha\frac{\partial\meanA_x}{\partial z}
+\frac{\partial^2\meanA_y}{\partial z^2},
\label{eq2}
\end{equation}
and $\meanA_z=0$.
In the following, all quantities are dimensionless.
We consider perfect conductor boundary condition on one side of the
domain ($z=0$).
This means that the electric field in the $xy$ plane vanishes on
the boundary.
Owing to the use of the Weyl gauge, the electrostatic potential gradient
is absent in \Eq{induct}, so the perfect conductor condition
implies that $\meanA_x=\meanA_y=0$.

On the other side of the domain, we assume a vacuum boundary condition.
For our one-dimensional domain, this means that $\meanB_x=\meanB_y=0$
\citep{RSS88}, which corresponds to $\partial_z\meanA_x=\partial_z\meanA_y=0$.
The most slowly decaying mode is a quarter sine wave, that is, $\meanA_x$
or $\meanA_y$ are proportional to $\sin z$ in $0\leq z\leq\pi/2$ \citep{BCC09}.

\section{Complex notation and integral constraints}

The basic approach used here is similar to that in other problems with
constant coefficients and in finite domains with boundary conditions, such
as the no-slip condition in Rayleigh--B\'enard convection \citep{Cha61}
or the pole-equator boundary conditions in $\alpha\Omega$ dynamos
\citep{Par71}.
Unlike convection, which is non-oscillatory at onset, we allow
here for oscillatory solutions.
Furthermore, we combine \Eqs{eq1}{eq2} into a single equation for
the complex variable
\EQ
{\cal A}\equiv\meanA_x+\ii\meanA_y.
\EN
Thus, \Eqs{eq1}{eq2} can be written as
\EQ
\frac{\partial{\cal A}}{\partial t}=\ii\alpha\frac{\partial{\cal A}}{\partial z}
+\frac{\partial^2{\cal A}}{\partial z^2}.
\EN
We now assume the solution to be of the form
\EQ
{\cal A}(z,t)=\hat{\cal A}(z)\,e^{-\ii\omega t},
\EN
where $\hat{\cal A}(z)$ obeys the ordinary differential equation
\EQ
\hat{\cal A}''+\ii\alpha\hat{\cal A}'+\ii\omega\hat{\cal A}=0,
\label{ode}
\EN
where primes denote $z$ derivatives.
The boundary conditions are
\EQ
\hat{\cal A}=0\quad\mbox{on $\;z=0$},
\label{bc1}
\EN
\EQ
\hat{\cal A}'=0\quad\mbox{on $\;z=\pi/2$}.
\label{bc2}
\EN
In general, $\omega$ can be complex, but since we are here interested
in marginally excited dynamos, we restrict ourselves in the following
to $\omega$ being real.

We now also assume that $\alpha$ is constant.
In that case, oscillatory solutions were previously thought impossible
\citep{RB87}.
Analogously to their approach, we multiply \Eq{ode} by $\hat{\cal A}^\ast$,
where the asterisk denotes complex conjugation, and integrate by parts.
Using \Eqs{bc1}{bc2}, we obtain
\begin{equation}
\int_0^{\pi/2}\hat{\cal A}''\hat{\cal A}^\ast\,\dd z
=-\int_0^{\pi/2}\left|\hat{\cal A}'\right|^2\,\dd z.
\end{equation}
Furthermore, $(\hat{\cal A}\hat{\cal A}^\ast)'
=\hat{\cal A}'\hat{\cal A}^\ast+\hat{\cal A}\hat{\cal A}'^\ast
=2\,\Rey(\hat{\cal A}'\hat{\cal A}^\ast)$, so
\begin{equation}
\hat{\cal A}'\hat{\cal A}^\ast
=\left(\half\left|\hat{\cal A}\,\right|^2\right)'
+\ii\,\Imag(\hat{\cal A}'\hat{\cal A}^\ast).
\end{equation}
\EEq{ode} yields altogether four terms, two of which are real and
the other two imaginary.
We obtain two integral constraints
\begin{equation}
\alpha=-\left.\int_0^{\pi/2}\left|\hat{\cal A}'\right|^2\,\dd z\;\right/
\int_0^{\pi/2}\Imag(\hat{\cal A}'\hat{\cal A}^\ast)\,\dd z,
\label{constraint1}
\end{equation}
\begin{equation}
\omega=-\half\alpha\,\left|\hat{\cal A}\,\right|^2_{\pi/2}\left/
\int_0^{\pi/2}\left|\hat{\cal A}\,\right|^2\,\dd z,\right.
\label{constraint2}
\end{equation}
where $|\hat{\cal A}\,|^2_{\pi/2}$ denotes the value of
$|\hat{\cal A}\,|^2$ on the second boundary at $z=\pi/2$.
This implies that $\alpha\omega\leq0$ (negative frequencies for positive
$\alpha$) and $\omega\neq0$ if $|\hat{\cal A}\,|_{\pi/2}>0$ and
$\alpha\neq0$.

Similar integral constraints can also be formulated for the complex
magnetic field, $\hat{\cal B}(z)=\ii\hat{\cal A}(z)$.
Unfortunately, the perfect conductor boundary condition,
$\ii\etaT\hat{\cal B}'=\alpha\hat{\cal B}$, is more cumbersome.
Instead, one could formulate the problem for an artificially
modified boundary condition, $\hat{\cal B}'=0$ on $z=0$.
Together with the condition $\hat{\cal B}=0$ on $z=\pi/2$, the problem
for $\hat{\cal B}(z)$ becomes equivalent to that for $\hat{\cal A}(z)$.
In either case, the integral constraints are analogous to those of
\cite{RB87}; see \App{IntegralConstraint} for details.

\section{The solution}

Given that \Eq{ode} has constant coefficients,
it has solutions proportional to
\EQ
\hat{\cal A}_i(z)\propto e^{\ii k_iz},
\label{solgen}
\EN
where the index $i$ denotes one of two independent solutions.
The $k_i$ are in general complex and obey the characteristic equation
\EQ
k^2+\alpha k-\ii\omega=0.
\EN
It has two solutions,
\EQ
k_\pm=-\alpha/2\pm\sqrt{\alpha^2/4+\ii\omega}.
\EN

\begin{figure}[t!]\begin{center}
\includegraphics[width=\columnwidth]{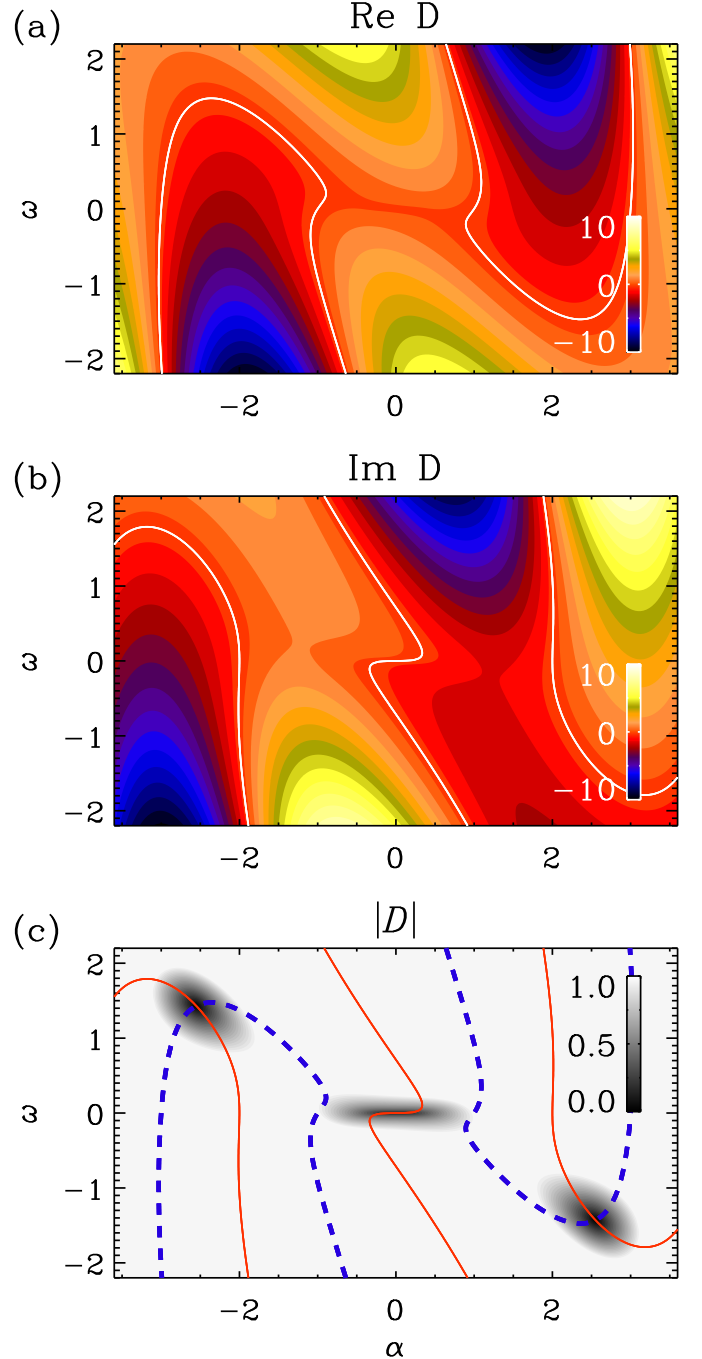}
\end{center}\caption[]{
Plots of (a) real, (b) imaginary, and (c) absolute parts of
$D(\alpha,\omega)$.
In (a) and (b), the zero lines are marked in white, while in (c) those
of $\Rey D$ are dotted blue and those of $\Imag D$ are solid red.
}\label{pcomplex}\end{figure}

To satisfy the boundary conditions \eqs{bc1}{bc2}, we write the solution
as a superposition of $e^{\ii k_+z}$ and $e^{\ii k_-z}$.
\EEq{bc1} is readily satisfied by writing
\EQ
\hat{\cal A}(z)=e^{\ii k_+z}-e^{\ii k_-z},
\label{sol}
\EN
where we have ignored the possibility of an arbitrary (complex) constant
in front of $\hat{\cal A}$.
To satisfy \Eq{bc2}, we now require that
\EQ
D(\alpha,\omega)=k_+ e^{\ii k_+\pi/2}-k_-e^{\ii k_-\pi/2}
\label{disc}
\EN
vanishes.
The existence of solutions to $D(\alpha,\omega)=0$ is
demonstrated by looking at a contour plot of $|D|$; see \Fig{pcomplex},
where we also plot separately the real and imaginary parts of $D$.
We see two zeroes in $D(\alpha,\omega)$, which is confirmed by the crossing
of the lines where $\Rey D$ and $\Imag D$ vanish.
[At $\alpha=\omega=0$, there is no such crossing, so $D(0)$ is not a solution.]
The transcendental equation relating $\alpha$ to $\omega$ can be written
in more explicit form as
\EQ
e^{\ii\pi\sqrt{\alpha^2/4+\ii\omega}}+\left.\left(\alpha/2
+\sqrt{\alpha^2/4+\ii\omega}\,\right)^2\right/(\ii\omega)=0.
\label{disc2}
\EN
To find solutions to $D(\alpha,\omega)=0$,
it is convenient to introduce the complex variable
\begin{equation}
Z\equiv\alpha+\ii\omega.
\end{equation}
We seek solutions to $D(Z)=0$ via complex interpolation,
\begin{equation}
Z=Z_0-D_0\,(Z_0-Z_{-1})/(D_0-D_{-1}),
\end{equation}
where subscripts $0$ and $-1$ refer to the current
and previous iteration.
This yields the first critical value as
\begin{equation}
Z_\ast=\alpha+\ii\omega\approx2.5506504-1.4296921\,\ii,
\label{eigenval}
\end{equation}
with the corresponding complex wavenumbers
\begin{equation}
k_+\approx0.10161896-0.51915398\,\ii,
\end{equation}
\begin{equation}
k_-\approx-2.6522693+0.51915398\,\ii.
\end{equation}
The wavenumbers $k_+$ and $k_-$ obey the relation
\EQ
k_++k_-+\alpha=0,
\EN
which follows from \Eqs{bc1}{sol}.
The critical values of $\alpha$ and $\omega$ were first obtained
by \cite{JBKR16} using explicit time integration.

\begin{figure}[t!]\begin{center}
\includegraphics[width=\columnwidth]{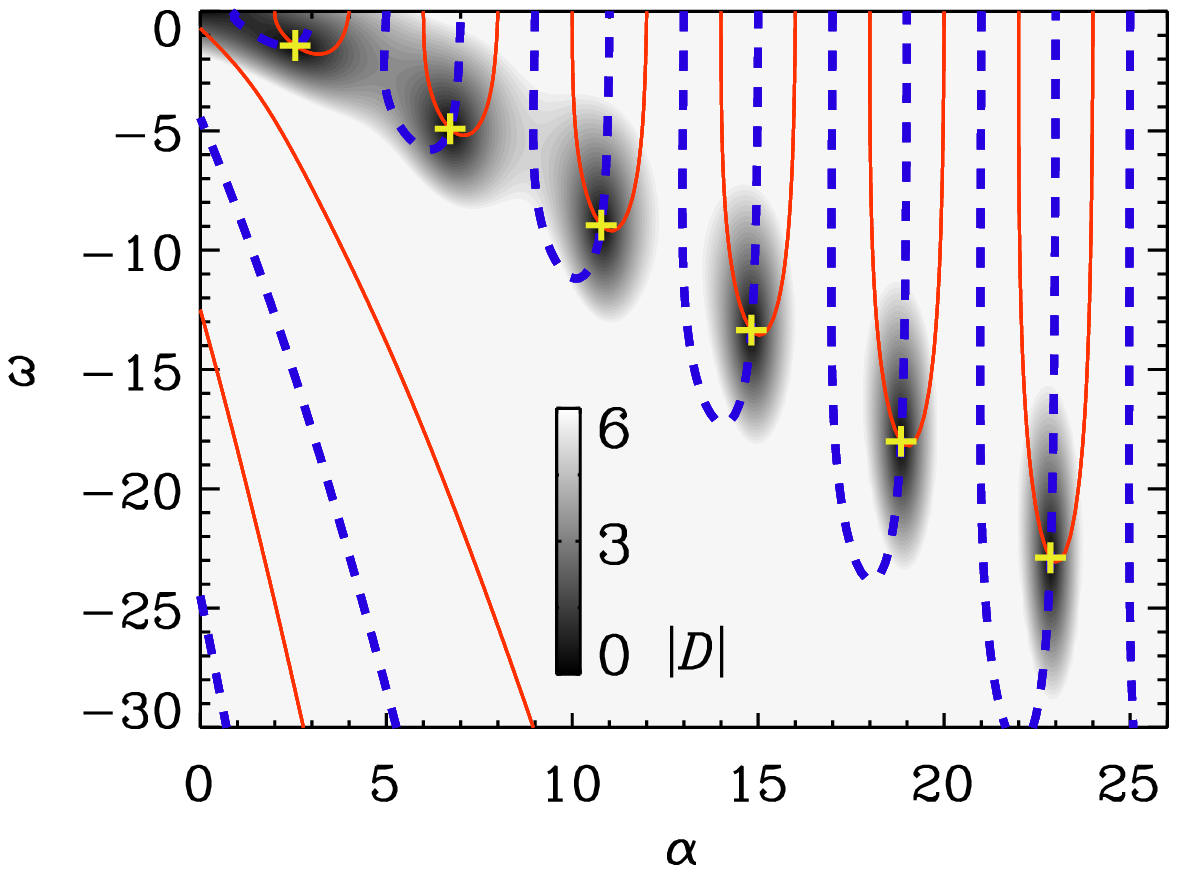}
\end{center}\caption[]{
Similar to \Figp{pcomplex}{c}, but for the next higher modes
($+$ signs).
}\label{pcomplex_more}\end{figure}

Additional solutions exist in the second and fourth quadrant of
the $\alpha\omega$ plane; see \Fig{pcomplex_more}.
They are all oscillatory, in agreement with the integral constraints;
see \Eqs{constraint1}{constraint2} and \Tab{THigherModes}.
However, those higher modes would generally be unstable in a nonlinear
calculation and therefore only of limited interest \citep{BKMMT89}.

\begin{table}[b!]\caption{
Critical values of $\alpha$ and $\omega$ for the higher modes.
}\vspace{12pt}\centerline{\begin{tabular}{ccc}
mode & $\alpha$ & $\omega$ \\
\hline
1 & 2.5506504 & $-1.4296921$ \\
2 & 6.7152255 & $-4.9166082$ \\
3 & 10.779288 & $-8.9553785$ \\
4 & 14.815829 & $-13.351365$ \\
5 & 18.840111 & $-18.013101$ \\
6 & 22.857683 & $-22.886942$ \\
7 & 26.871119 & $-27.937488$ \\
8 & 30.881799 & $-33.139583$ \\
\label{THigherModes}\end{tabular}}\end{table}

\begin{figure}[t!]\begin{center}
\includegraphics[width=\columnwidth]{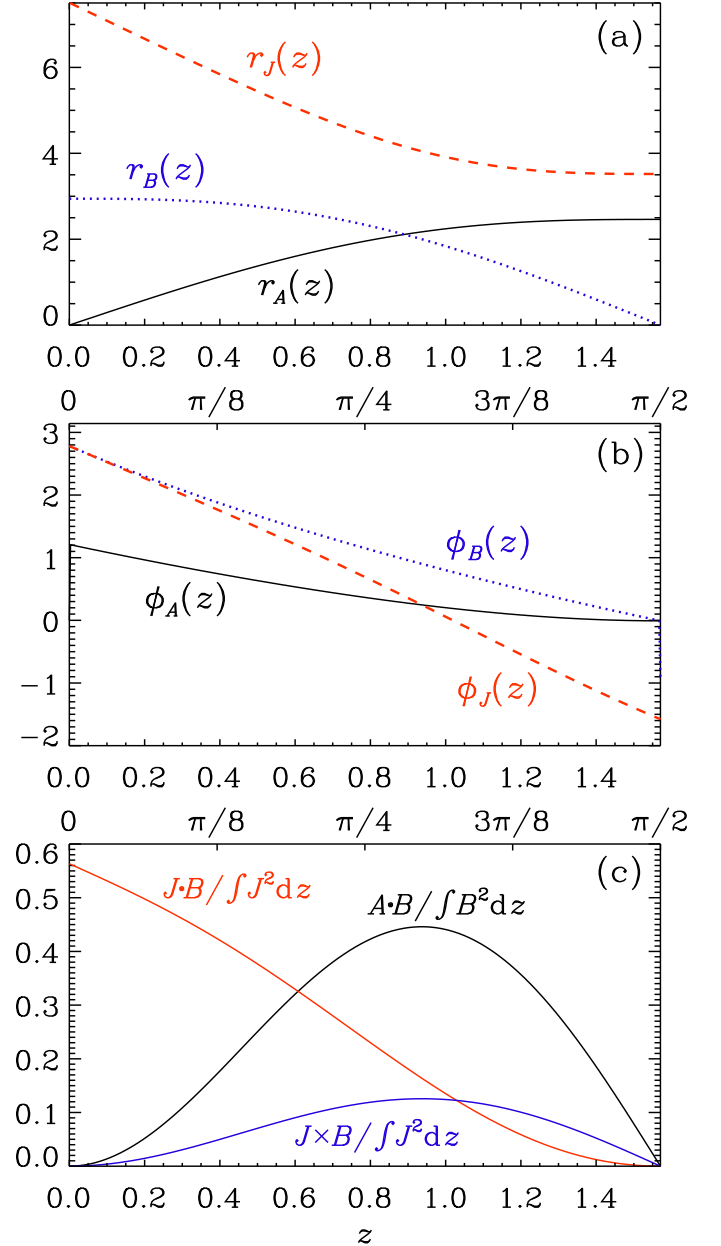}
\end{center}\caption[]{
(a) Moduli and (b) phases of $\hat{\cal A}(z)$, $\hat{\cal B}(z)$,
and $\hat{\cal J}(z)$, as well as (c) normalized magnetic and current
helicity densities together with the $z$ component of the Lorentz force.
}\label{pplot2}\end{figure}

The solution is now completely described by the value of $Z_\ast$.
It is convenient to write the solution in the form
\begin{equation}
\hat{\cal A}=r_A(z) \, e^{\ii\phi_A(z)},
\end{equation}
where $r_A(z)$ and $\phi_A(z)$ are amplitude and phase of $\hat{\cal A}$.
In view of computing magnetic field and current density, we also define
\begin{equation}
\hat{\cal B}\equiv \ii\hat{\cal A}'=r_B(z) \, e^{\ii\phi_B(z)}
\end{equation}
and
\begin{equation}
\hat{\cal J}\equiv -\hat{\cal A}''=r_J(z) \, e^{\ii\phi_J(z)},
\end{equation}
respectively.
In \Fig{pplot2} we plot the moduli and phases of $\hat{\cal A}(z)$,
$\hat{\cal B}(z)$, and  $\hat{\cal J}(z)$.
Note that $r_A(0)=0$, as required by \Eq{bc1}, and
$r_A'(\pi/2)=\phi_A'(\pi/2)=0$, as required by \Eq{bc2}.
In general, however, $\hat{\cal J}(0)\equiv-\hat{\cal A}''(0)\neq0$.
The derivative of the phase is an ``effective'' wavenumber,
$k^{(B)}_{\rm eff}=\dd\phi_B/\dd z$, and determines the $z$-dependent
phase speed $c=\omega/k^{(B)}_{\rm eff}$, which is positive for
positive $\alpha$, so the wave moves in the positive $z$ direction.

In \Figp{pplot2}{c} we plot the magnetic and current helicity densities,
as well as the $z$ component of the Lorentz force,
\EQ
\meanAA\cdot\meanBB=\Rey\hat{\cal A}^\ast\hat{\cal B},\quad
\meanJJ\cdot\meanBB=\Rey\hat{\cal J}^\ast\hat{\cal B},\quad
(\meanJJ\times\meanBB)_z=\Imag\hat{\cal J}^\ast\hat{\cal B},\quad
\EN
normalized by $\int\meanBB^2\dd z\equiv\int|\hat{\cal B}\,|^2\dd z$
and $\int\meanJJ^2\dd z\equiv\int|\hat{\cal J}\,|^2\dd z$ for the first,
and second and third quantities, respectively.
The Lorentz force has a maximum at $z=0.937$, which is also the point
where the magnetic helicity density in the Weyl gauge has a maximum.
The current helicity density, however, has a maximum at $z=0$.
The ratio between the integrals of the two helicity densities is
\EQ
k_{\rm m}^2\equiv\left.\int\Rey\hat{\cal J}^\ast\hat{\cal B}\dd z\;\right/
\int\Rey\hat{\cal A}^\ast\hat{\cal B}\dd z,
\EN
where $k_{\rm m}$ denotes the wavenumber of the mean field; see Eq.~(25)
of \cite{BB02}.
For $\alpha^2$ dynamos in periodic domains, one finds $k_{\rm m}/k_1=1$,
but here we obtain $k_{\rm m}/k_1\approx2.253027$.
Interestingly, this is also the value of the magnetic Taylor microscale
wavenumber of the mean field, $k_{\rm T}$, defined through
$k_{\rm T}^2=\left.\int|\hat{\cal J}\,|^2\dd z\;\right/\!\int|\hat{\cal B}\,|^2\dd z$,
i.e., $k_{\rm T}=k_{\rm m}$.
Finally, for the fractional current helicity of the mean field \citep{BB02},
\EQ
\epsilon_{\rm m}=\left.\int\Rey\hat{\cal J}^\ast\hat{\cal B}\,\dd z\;\right/\,
\left(\int|\hat{\cal J}\,|^2\dd z\int|\hat{\cal B}\,|^2\dd z\right)^{1/2},
\EN
we find $\epsilon_{\rm m}\approx0.883315$, which is close to the value
$\epsilon_{\rm m}=1$ for $\alpha^2$ dynamos in periodic domains \citep{BB02}.

\begin{figure}[t!]\begin{center}
\includegraphics[width=\columnwidth]{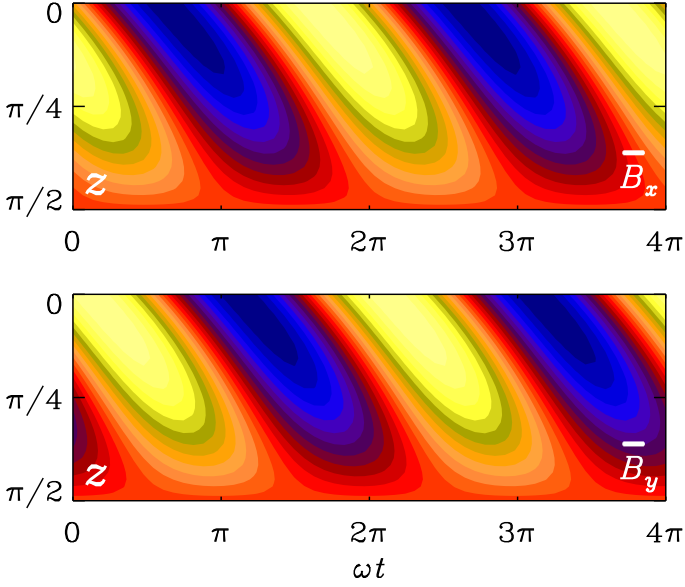}
\end{center}\caption[]{
Butterfly diagrams for $\meanB_x$ and $\meanB_y$,
with $z$ increasing downward.
}\label{pcontour2}\end{figure}

To plot butterfly diagrams of $\meanB_x$ and $\meanB_y$, we can now
write the fully time-dependent magnetic field as
\begin{equation}
\renewcommand*{\arraystretch}{1.3}
\begin{array}{llll}\itemsep5pt
\meanB_x(z,t)=r_B(z) \cos[\phi_B(z)-\omega t],\cr
\meanB_y(z,t)=r_B(z) \sin[\phi_B(z)-\omega t].
\end{array}
\end{equation}
This also demonstrates that the magnetic energy density,
\begin{equation}
E_{\rm M}=\half\meanBB^2=\half r_B(z)^2=E_{\rm M}(z),
\end{equation}
is independent of time and only a function of $z$.
In fact, the magnetic and current helicity densities,
as well as the $z$ component of the Lorentz force, 
all shown in \Figp{pplot2}{c}, are also independent of time.
The results for $\meanB_x(z,t)$ and $\meanB_y(z,t)$ are shown
in \Fig{pcontour2}, where $z$ increases downward so as to facilitate
comparison with Fig.~2 of \cite{BCC09}, who adopted a perfect conductor
boundary condition at high latitudes and a vacuum condition at the
equator.
In their case, however, $\alpha$ was non-constant and vanishing
on the equator.

\section{Discussion}

The graphs of the solutions obtained here look rather simple, but
would have been impossible to guess based on previous experience with
one-dimensional dynamos with vacuum field conditions on both ends of
the domain.
The field components of those dynamos are proportional to
$\cos z \, e^{\ii z}$.
Such dynamos have been studied extensively in connection with
demonstrating the asymptotically equal growth rates of even and odd
dynamo modes \citep{BKMMT89}, the behavior of dynamos in the highly
nonlinear regime \citep{MB90}, and the effects of magnetic helicity fluxes
\citep{BD01}.
Thus, one might have expected that the solution to the present problem
could have been expanded in terms of sine functions proportional to
$\sin\,(2n+1)z$ with integers $n\ge0$.
Such functions obey the boundary conditions of $\meanA_x$ on $z=0$ and $\pi/2$.
However, one sees immediately that such a solution for $\meanA_x$ would imply
that $\meanA_y$ has terms proportional to $\cos\,(2n+1)z$, which would then
violate the boundary conditions on $\meanA_y$ on both boundaries;
see \App{Expansion} for details.
This is indeed be a problem for spectral codes that employ sine or cosine
transforms; see \cite{VBJ08a,VBJ08b} for detailed studies and alternative
approaches.
It can also be a problem for codes that use symmetry conditions to populate
the ghost zones outside the computational mesh, as is done by default in
the {\sc Pencil Code}\footnote{\url{https://github.com/pencil-code}}.
This highlights once more the significance of having an independent and
analytic solution of such a dynamo.
To demonstrate this, we summarize in \Tab{TSum} the values of $\alpha$
and $|\omega|$ for a marginally excited dynamo obtained by using
either one-sided ({\tt 1s}) finite difference formulae on the boundaries or
symmetry/antisymmetry ({\tt s}) conditions \citep{Bra03} for different
meshpoint numbers $N_{\rm mesh}$.
The {\tt 1s} scheme does not restrict the second derivative and is found
to be slightly better than the {\tt s} scheme.

\begin{table}[t!]\caption{
Values of $\alpha$ and $|\omega|$ using one-sided ({\tt 1s})
finite difference formulae on the boundaries and symmetry/antisymmetry
({\tt s}) conditions for different meshpoint numbers $N_{\rm mesh}$.
Agreement with the analytic solution (``exact'') is indicated in bold face.
}\vspace{12pt}\centerline{\begin{tabular}{ccccc}
$N_{\rm mesh}$ &
$\alpha^{~\tt(1s)}$ & $|\omega|^{~\tt(1s)}$ &
$\alpha^{~\tt(s)}$  & $|\omega|^{~\tt(s)}$ \\
\hline
~~32&    2.55213 &    1.4350 &    2.55228 &    1.4289 \\  
128 &    2.55071 &    1.4298 &    2.55074 &    1.4297 \\  
512 &{\bf2.55065}&{\bf1.4297}&{\bf2.55065}&{\bf1.4297}\\  
exact&   2.55065 &    1.4297\\
\label{TSum}\end{tabular}}\end{table}

We have here also been able to find higher order modes.
They all lie in the same two quadrants in the $\alpha\omega$ plane.
Thus, for positive $\alpha$, we always have $\omega<0$.
When determining $\omega$ empirically from the period of the
oscillation, it would not have a definite sign, although
the sign has implications for the phase speed.
For $\alpha\Omega$ dynamos with differential rotation gradient $\Omega'$
in periodic domains with real wavenumber $k$, self-excited solutions
exist only when $\sgn[(k\alpha\Omega')\omega]>0$; see \App{AppA} and
Table~3 of \cite{BS05}.
However, unlike $\alpha\Omega$ dynamos, where both migration
directions are possible, depending just on $\sgn(\alpha\Omega')$,
for oscillatory $\alpha^2$ dynamos, the migration direction
is always away from the (perfect) conductor toward the vacuum.
This agrees with earlier findings for oscillatory $\alpha^2$
dynamos with nonuniform $\alpha$ profiles \citep{BCC09}.

In the context of oscillatory $\alpha\Omega$ dynamos, boundary conditions
have long been known to introduce behaviors that are not obtained for
infinite domains \citep{Par71}.
The antisymmetry condition at the equator was found to play the role of an
absorbing boundary that led to localized wall modes \citep{WKTP97,TPK97}.
Subsequent work using complex amplitude equations for the envelope of a
wave train demonstrated that boundary conditions can play a decisive role
in determining the migration direction of traveling waves \citep{TPK98}.
They emphasized that the traveling wave behavior is linked to the
symmetry-breaking in the mean-field dynamo equations.
This rather general result could explain the migration direction
of the $\alpha^2$ dynamo studied here.
The symmetry breaking, which occurs here through the boundary conditions,
might also be responsible for the occurrence of an oscillatory mode rather
than the non-selfadjointness mentioned in the introduction \citep{SSR85}.

\section{Conclusions}

The present work has shown that $\alpha^2$ dynamos with constant $\alpha$
can have oscillatory solutions provided the boundary conditions
on the two ends of the domain are different.
It is possible to construct a one-dimensional analytic solution characterized
by a complex function $\hat{\cal A}(z)$, which obeys Dirichlet and
von Neumann boundary conditions on the two ends of the domain.
The solution has been obtained as a superposition of two harmonic
functions with complex wavenumbers.
In principle, we could have solved the problem directly for
$\hat{\cal B}(z)=\ii\hat{\cal A}(z)$, but the boundary condition on $z=0$,
namely $\ii\etaT\hat{\cal B}'=\alpha\hat{\cal B}$, would be more complicated.
Integral constraints on $\hat{\cal B}$ would then be harder to impose,
unless one changed the perfect conductor boundary condition to
$\hat{\cal B}'=0$.
In that case, the problem becomes equivalent to the one considered
here if we replace $\hat{\cal A}\to\hat{\cal B}$.
In this connection, it should be noted that the very assumption
of a finite $\alpha$ effect on a perfect conductor boundary, while
mathematically sound, is physically not strictly realistic, because an
impenetrable boundary would necessarily make $\alpha$ anisotropic such
that its tangential components would vanish \citep{Rae82}.
Nevertheless, various DNS with helically forced turbulence extending
all the way to the walls confirm the presence of oscillatory migratory
solutions \citep{MTKB10,WBM11,JBKR16}.

Owing to our restriction to Cartesian geometry, the main application of
this model lies in the comparison with other numerical solutions in the same
geometry \citep[see, e.g.,][]{JBKR16}.
The present solution demonstrates clearly that a model with constant $\alpha$
is possible and has time-independent magnetic energy density.
Thus, when looking only at the rms value of the magnetic field or
the volume-integrated energy, one will not notice the presence of an
oscillatory solution.

When the $\alpha^2$ dynamo is applied to a star, $\alpha$ would have the
opposite sign on the other side of the equator (here for $z>\pi/2$) and
would then be described by a step function.
In that case, the field could be either symmetric or antisymmetric
about the equator.
Earlier work with a linear $\alpha$ profile suggests that the
antisymmetric solution is more easily excited \citep{BCC09,CBKK16}.
Such solutions would have a discontinuity in the derivative of the
current density at the equator.
More dramatic, however, would be the case of symmetric solutions when
a vacuum or vertical field condition is assumed on the outer boundary,
because in that case the current density itself would be discontinuous
at the equator.
Interestingly, the critical values of $\alpha$ are the same in both cases.
While a step function profile of $\alpha$ is artificial,
it does pose a simple benchmark for numerical schemes.
The analytic solution presented here applies also to this case.
This solution may also serve as a pedagogical illustration that
oscillatory migratory dynamos with constant $\alpha$ are possible.

\begin{acknowledgements}
I thank Ben Brown, Matthias Rheinhardt, and an anonymous referee
for useful remarks.
This work was supported in part by
the Swedish Research Council grant No.\ 2012-5797,
and the Research Council of Norway under the FRINATEK grant 231444.
This work utilized the Janus supercomputer, which is supported by the
National Science Foundation (award number CNS-0821794), the University
of Colorado Boulder, the University of Colorado Denver, and the National
Center for Atmospheric Research. The Janus supercomputer is operated by
the University of Colorado Boulder.
\end{acknowledgements}

\appendix

\section{Integral constraint in multi-dimensions}
\label{IntegralConstraint}

The purpose of this appendix is to demonstrate the analogy between
\Eqs{constraint1}{constraint2} and the corresponding one of \cite{RB87}.
However, instead of assuming the dynamo region to be surrounded by
vacuum and extending some of the volume integrals over all space,
we adopt here perfect conductor and vertical field boundary conditions.
In a multi-dimensional domain, the latter is no longer a proper vacuum condition,
but it can be motivated as being a more realistic representation of
stellar surface fields affected by magnetic buoyancy effects \citep{Yos75}.
Multiplying by $\hatBB^\ast$, the dynamo eigenvalue problem takes the form
\begin{equation}
-\hatBB^\ast\cdot(\nab\times\nab\times\hatBB)
+\hatBB^\ast\cdot\nab\times(\alpha\hatBB)+\ii\omega|\hatBB|^2=0.
\end{equation}
Using
\begin{equation}
2\ii\alpha\,\Imag\left(\hatBB^\ast\cdot\nab\times\hatBB\right)
=\nab\cdot\left(\alpha\hatBB\times\hatBB^\ast\right)
-\nab\alpha\cdot\left(\hatBB\times\hatBB^\ast\right),\quad
\end{equation}
but assuming now constant $\alpha$ in a volume $V$, we obtain
\begin{equation}
\alpha=-\left.\int_V\left|\nab\times\hatBB\right|^2\,\dd V\,\right/\!
\int_V\Imag\left(\hatBB\cdot\nab\times\hatBB^\ast\right)\,\dd V
\end{equation}
and, as in \cite{RB87},
\begin{equation}
\omega=-\half\alpha\left.\oint_{\partial V}
\Imag\left(\hatBB\times\hatBB^\ast\right)\cdot\dd\SSS\,
\right/\!\int_V\left|\hatBB\right|^2\dd V.
\label{RB87constraint}
\end{equation}
These equations are analogous to \Eqs{constraint1}{constraint2}.
By comparison, \cite{RB87} assumed a potential field on all boundaries,
so $\hatBB=-\nab\Phi$, where $\Phi$ is the magnetic scalar potential.
Writing the integrand of the surface integral in \Eq{RB87constraint}
as $\nab\times(\Phi\nab\Phi^\ast)$ and turning the surface integral
back into a volume integral, one sees that the divergence of the curl
vanishes, and therefore $\omega=0$.
However, this does not apply to our case where we have  different
boundary conditions on the two ends.
By comparison, in one-dimensional dynamos with vacuum conditions
on both ends, $|\hat{\cal A}|^2$ has, in a non-transient state
and with the gauge $\int\hat{\cal A}\,\dd z=0$, the same value on
both boundaries, so \Eq{constraint2} does indeed predict $\omega=0$.

\section{Quarter sine wave expansion}
\label{Expansion}

In this appendix we give the results for a quarter sine wave expansion
of $\hat{\cal A}$,
\EQ
\hat{\cal A}(z)=\sum_{n=0}^\infty \hat{\cal A}_n \sin\,(2n+1)z,
\EN
where each element of the expansion obeys \Eqs{bc1}{bc2}.
The coefficients are given by
$\hat{\cal A}_n=\int_{0}^{\pi/2} \hat{\cal A} \sin\,(2n+1)z$.
We have strictly $\hat{\cal A}''(0)=0$, although the analytic value
is nonvanishing, $\hat{\cal A}''(0)\approx7.0242061-2.6483598\,\ii$.
For $\hat{\cal A}'(0)$ we have
\EQ
\hat{\cal A}'(0)\to {\cal S}_N\equiv\sum_{n=0}^N(2n+1)\,\hat{\cal A}_{n},
\EN
which converges extremely slowly to the analytic value obtained from \Eq{sol},
which is $\hat{\cal A}'(0)\approx1.0383077+2.7538882\,\ii$; see \Tab{TConv},
where we list the first few values of ${\cal S}_n$ and $\hat{\cal A}_n$.

\begin{table}[t!]\caption{
Coefficients $\hat{\cal A}_n$ and ${\cal S}_n$.
}\vspace{12pt}\centerline{\begin{tabular}{rrrrr}
$n$ & $\Rey\,\hat{\cal A}_n$ & $\Imag\,\hat{\cal A}_n$ &
$\Rey\, {\cal S}_n$ & $\Imag\, {\cal S}_n$ \\
\hline
  0 & $ 2.512$ & $ 0.493$ & $ 2.512$ & $ 0.493$ \\
  1 & $-0.052$ & $ 0.557$ & $ 2.355$ & $ 2.165$ \\
  2 & $-0.114$ & $ 0.054$ & $ 1.788$ & $ 2.437$ \\
  3 & $-0.024$ & $ 0.013$ & $ 1.622$ & $ 2.527$ \\
  4 & $-0.015$ & $ 0.006$ & $ 1.486$ & $ 2.578$ \\
  5 & $-0.006$ & $ 0.003$ & $ 1.418$ & $ 2.609$ \\
  6 & $-0.005$ & $ 0.002$ & $ 1.358$ & $ 2.631$ \\
  8 & $-0.002$ & $ 0.001$ & $ 1.287$ & $ 2.659$ \\
 10 & $-0.001$ & $ 0.000$ & $ 1.242$ & $ 2.677$ \\
100 & $-0.000$ & $ 0.000$ & $ 1.060$ & $ 2.746$ \\
500 & $ 0.000$ & $ 0.000$ & $ 1.044$ & $ 2.752$ \\
\multicolumn{3}{r}{analytic solution $\longrightarrow$} & 1.038 & 2.753 \\
\label{TConv}\end{tabular}}\end{table}

\section{Comparison with the $\alpha\Omega$ dynamo}
\label{AppA}

The purpose of this appendix is to show that for $\alpha\Omega$ dynamos,
$\alpha\omega\Omega'k>0$ and $\alpha c\Omega'>0$,
where $c=\omega/k$ is the phase speed.
We assume a linear shear flow velocity $\meanUU=(0,x\Omega',0)$,
where $\Omega'$ is the velocity gradient.
Using the advective gauge, $\meanUU\cdot\meanAA=0$ \citep{BNST95,CHBM11},
we have
\begin{equation}
\frac{\partial\meanA_x}{\partial t}=-\Omega'\,\meanA_y
+\etaT\frac{\partial^2\meanA_x}{\partial z^2},
\label{eq1b}
\end{equation}
\begin{equation}
\frac{\partial\meanA_y}{\partial t}=+\alpha\frac{\partial\meanA_x}{\partial z}
+\etaT\frac{\partial^2\meanA_y}{\partial z^2}.
\label{eq2b}
\end{equation}
The dispersion relation is then
\begin{equation}
-\ii\omega\equiv-\ii kc=-\etaT k \pm (-\ii k\alpha\Omega')^{1/2}.
\end{equation}
Using $(2\,\ii)^{1/2}=1+\ii$ and $(-2\,\ii)^{1/2}=(1+\ii)\ii=-1+\ii$, we have
\begin{equation}
-\ii\omega\equiv-\ii kc=-\etaT k
\pm \left[\ii-\sgn(k\alpha\Omega')\right]
\,\left|k\alpha\Omega'/2\right|^{1/2}.
\end{equation}
For positive (negative) values of $k\alpha\Omega'$, only the
lower (upper) sign yields marginally excited dynamos, so
\begin{equation}
\sgn\omega = \sgn(k\alpha\Omega') \quad\mbox{and}\quad
\sgn c = \sgn(\alpha\Omega').
\end{equation}
Thus, the migration direction depends just on the sign of $\alpha\Omega'$,
but the frequency depends also on the sign of $k$.



\begin{thebibliography}{}

\bibitem[Baryshnikova \& Shukurov(1987)]{BS87}
Baryshnikova, Y. \& Shukurov, A. M.\yan{1987}{308}{89}

\bibitem[Blackman \& Brandenburg(2002)]{BB02}
Blackman, E. G., \& Brandenburg, A.\yapj{2002}{579}{359}

\bibitem[Brandenburg(2001)]{B01}
Brandenburg, A.\yapj{2001}{550}{824}

\bibitem[Brandenburg(2003)]{Bra03}
Brandenburg, A.\yproc{2003}{269}
{Advances in nonlinear dynamos
(The Fluid Mechanics of Astrophysics and Geophysics, Vol.\ {\bf9})}
{A. Ferriz-Mas \& M. N\'u\~nez}
{Taylor \& Francis, London and New York}

\bibitem[Brandenburg \& Dobler(2001)]{BD01}
Brandenburg, A., \& Dobler, W.\yana{2001}{369}{329}

\bibitem[Brandenburg \& Subramanian(2005)]{BS05}
Brandenburg, A., \& Subramanian, K.\yjour{2005}{Phys.\ Rep.}{417}{1}

\bibitem[Brandenburg et al.(2009)]{BCC09}
Brandenburg, A., Candelaresi, S., \& Chatterjee, P.\ymn{2009}{398}{1414}

\bibitem[Brandenburg et al.(1989)]{BKMMT89}
Brandenburg, A., Krause, F., Meinel, R., Moss, D., \& Tuominen, I.\yana{1989}{213}{411}

\bibitem[Brandenburg et al.(1995)]{BNST95}
Brandenburg, A., Nordlund, \AA., Stein, R. F., \& Torkelsson, U.\yapj{1995}{446}{741}

\bibitem[Candelaresi et al.(2011)]{CHBM11}
Candelaresi, S., Hubbard, A., Brandenburg, A., \& Mitra, D.\ypp{2011}{18}{012903}

\bibitem[Chandrasekhar(1961)]{Cha61}
Chandrasekhar, S.\ybook{1961}{Hydrodynamic and Hydromagnetic Stability}
{Dover Publications, New York}, Sect.~15, pp.\ 37

\bibitem[Cole et al.(2016)]{CBKK16}
Cole, E., Brandenburg, A., K\"apyl\"a, P. J., \& K\"apyl\"a, M. J.\yana{2016}{593}{A134}


\bibitem[Jabbari et al.(2016)]{JBMKR16}
Jabbari, S., Brandenburg, A., Mitra, D., Kleeorin, N., \& Rogachevskii, I.\ymn{2016}{459}{4046}

\bibitem[Jabbari et al.(2017)]{JBKR16}
Jabbari, S., Brandenburg, A., Kleeorin, N., \& Rogachevskii, I.\dmn{2017}
{10.1093/mnras/stx148}

\bibitem[K\"apyl\"a et al.(2013)]{KMCWB13}
K\"apyl\"a, P. J., Mantere, M. J., Cole, E., Warnecke, J., \& Brandenburg, A.\yapj{2013}{778}{41}

\bibitem[Masada \& Sano(2014)]{MS14}
Masada, Y., \& Sano, T.\yapjl{2014}{794}{L6}

\bibitem[Maunder(1904)]{Mau04}
Maunder, E. W.\ymn{1904}{64}{747}

\bibitem[Meinel \& Brandenburg(1990)]{MB90}
Meinel, R., \& Brandenburg, A.\yana{1990}{238}{369}

\bibitem[Mitra et al.(2009)]{MTBM09}
Mitra, D., Tavakol, R., Brandenburg, A., \& Moss, D.\yapj{2009}{697}{923}

\bibitem[Mitra et al.(2010)]{MTKB10}
Mitra, D., Tavakol, R., K\"apyl\"a, P. J., \& Brandenburg, A.\yapjl{2010}{719}{L1}

\bibitem[Mitra et al.(2014)]{MBKR14}
Mitra, D., Brandenburg, A., Kleeorin, N., \& Rogachevskii, I.\ymn{2014}{445}{761}

\bibitem[Parker(1955)]{Par55}
Parker, E. N.\yapj{1955}{122}{293}

\bibitem[Parker(1971)]{Par71}
Parker, E. N.\yapj{1971}{164}{491}

\bibitem[Pouquet et al.(1976)]{PFL76}
Pouquet, A., Frisch, U., \& L\'eorat, J.\yjfm{1976}{77}{321}

\bibitem[R\"adler(1980)]{Rae80}
R\"adler, K.-H.\yana{1980}{301}{101}

\bibitem[R\"adler(1982)]{Rae82}
R\"adler, K.-H.\ygafd{1982}{20}{191}

\bibitem[R\"adler \& Br\"auer(1987)]{RB87}
R\"adler, K.-H., \& Br\"auer, H.-J.\yan{1987}{308}{101}

\bibitem[Ruzmaikin et al.(1980)]{RSS80}
Ruzmaikin, A. A., Sokolov, D. D., \& Shukurov, A. M.\ymhdn{1980}{16}{15}

\bibitem[Ruzmaikin et al.(1988)]{RSS88}
Ruzmaikin, A. A., Sokoloff, D. D. \& Shukurov, A. M.\ybook{1988}
{Magnetic Fields of Galaxies}{Kluwer, Dordrecht}

\bibitem[Shukurov et al.(1985)]{SSR85}
Shukurov, A. M., Sokolov, D. D., \& Ruzmaikin, A. A.\ymhdn{1985}{3}{6};
translated from \yjour{1985}{Magnitnaia Gidrodinamika}{3}{9}

\bibitem[Steenbeck \& Krause(1969a)]{SK69a}
Steenbeck, M., \& Krause, F.\yan{1969a}{291}{49}

\bibitem[Steenbeck \& Krause(1969b)]{SK69b}
Steenbeck, M., \& Krause, F.\yan{1969b}{291}{271}

\bibitem[Steenbeck et al.(1966)]{SKR66}
Steenbeck, M., Krause, F., \& R\"adler, K.-H.\yjour{1966}{Z. Naturforsch.}{21a}{369}

\bibitem[Stefani \& Gerbeth(2003)]{SG03}
Stefani, F., \& Gerbeth, G.\ypre{2003}{67}{027302}

\bibitem[Tobias et al.(1997)]{TPK97}
Tobias, S. M., Proctor, M. R. E., \& Knobloch, E.\yana{1997}{318}{L55}

\bibitem[Tobias et al.(1998)]{TPK98}
Tobias, S. M., Proctor, M. R. E., \& Knobloch, E.\yphyd{1998}{113}{43}

\bibitem[Vasil et al.(2008a)]{VBJ08a}
Vasil, G. M., Brummell, N. H., \& Julien, K.\yjcp{2008a}{227}{7999}

\bibitem[Vasil et al.(2008b)]{VBJ08b}
Vasil, G. M., Brummell, N. H., \& Julien, K.\yjcp{2008b}{227}{8017}

\bibitem[Warnecke et al.(2011)]{WBM11}
Warnecke, J., Brandenburg, A., \& Mitra, D.\yana{2011}{534}{A11}

\bibitem[Warnecke et al.(2014)]{WKKB14}
Warnecke, J., K\"apyl\"a, P. J., K\"apyl\"a, M. J., \& Brandenburg, A.\yapjl{2014}{796}{L12}

\bibitem[Vainshtein \& Ruzmaikin(1971)]{VR71}
Vainshtein, S. I., \& Ruzmaikin, A. A.\ysov{1971}{16}{365}

\bibitem[Worledge et al.(1997)]{WKTP97}
Worledge, D., Knobloch, E., Tobias, S., \& Proctor, M.\yprsa{1997}{453}{119}

\bibitem[Yoshimura(1975)]{Yos75}
Yoshimura, H.\yapjs{1975}{29}{467}

\end{thebibliography}
\end{document}